%% file: main.tex
\documentclass{sig-alternate}
\newcommand{\ignore}[1]{}
\usepackage[pass]{geometry}
\usepackage{fancyhdr}
\usepackage[normalem]{ulem}
\usepackage[hyphens]{url}
\usepackage{color}
\usepackage{soul}
\usepackage{colortbl}

\usepackage[bookmarks=true,breaklinks=true,letterpaper=true,colorlinks,linkcolor=black,citecolor=blue,urlcolor=black]{hyperref}

\fancypagestyle{firstpage}{
	\fancyhf{}
	\setlength{\headheight}{50pt}
	
	\fancyhead[C]{
			} 
	\pagenumbering{arabic}
}

\title{Boosting LSTM Performance Through Dynamic Precision Selection}

\author{Franyell Silfa, Jose-Maria Arnau, Antonio Gonz\`alez\\
Computer Architecture Deparment, Universitat Politecnica de Catalunya\\
\{fsilfa, jarnau, antonio\}@ac.upc.edu
}

\begin{document}
\maketitle
\thispagestyle{firstpage}
\pagestyle{plain}


\input{abstract}

\input{introduction}

\input{background}

\input{techniques}

\input{methodology}

\input{results}

\input{related_work}

\input{conclusions}

\newpage


\bibliographystyle{ieeetr}
\bibliography{ref}

\end{document}

%% file: abstract.tex
\begin{abstract}
The use of low numerical precision is a fundamental optimization included in modern accelerators for Deep Neural Networks (DNNs). The number of bits of the numerical representation is set to the minimum precision that is able to retain accuracy based on an offline profiling, and it is kept constant for DNN inference. 

In this work, we explore the use of dynamic precision selection during DNN inference. We focus on Long Short Term Memory (LSTM) networks, which represent the state-of-the-art networks for applications such as machine translation and speech recognition. Unlike conventional DNNs, LSTM networks remember information from previous evaluations by storing data in the LSTM cell state. Our key observation is that the cell state determines the amount of precision required: time steps where the cell state changes significantly require higher precision, whereas time steps where the cell state is stable can be computed with lower precision without any loss in accuracy.

Based on this observation, we implement a novel hardware scheme that tracks the evolution of the elements in the LSTM cell state and dynamically selects the appropriate precision in each time step. For a set of popular LSTM networks, our scheme selects the lowest precision for more than 66\% of the time, outperforming systems that fix the precision statically. We evaluate our proposal on top of a modern accelerator highly optimized for LSTM computation, and show that it provides 1.56x speedup and 23\% energy savings on average without any loss in accuracy. The extra hardware to determine the appropriate precision represents a small area overhead of 8.8\%.
\end{abstract}

%% file: introduction.tex
\section{Introduction}\label{s:introduction}

\begin{figure}[t!]
	\centering
	\includegraphics[width=3.375in]{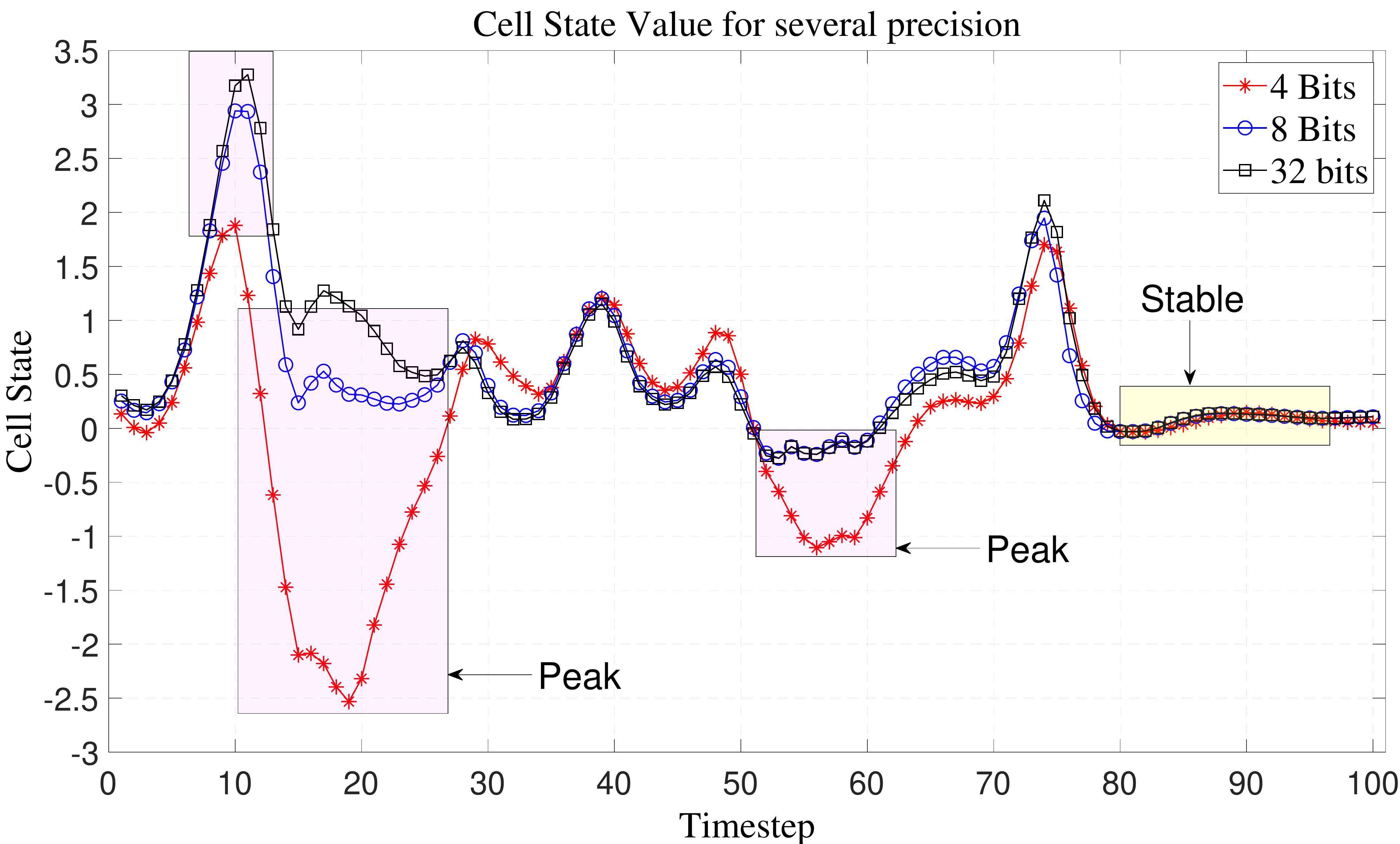}
	\caption{Evolution of one element in the cell state of a speech recognition LSTM network~\cite{wu2016google}. For stable regions the low precision (4 bit version) evaluation accurately tracks the behavior of the high precision (FP32) version. However, a large error is introduced when the tracked element is on a peak.}
	\label{f:cellstate_peak_no_fix}
\end{figure}

Long Short Term Memory (LSTM) neural networks represent the state-of-the-art solution for sequence-to-sequence problems such as machine translation~\cite{wu2016google}, automatic caption generation~\cite{Vinyals_2015_CVPR} or speech recognition~\cite{gravesSpeech2013}. Unlike conventional DNNs, LSTMs store information from previous executions to improve the accuracy of future prediction. In addition, they can handle input and output sequences of variable length. However, their recurrent nature severely constrains the amount of parallelism, making it challenging to achieve low latency LSTM inference on CPUs~\cite{deepcpu2018} and GPUs~\cite{appleyard2016optimizing}. Not surprisingly, accelerators to boost LSTM performance have been recently presented~\cite{jouppi2017TPU,fsilfa2018,guan2017fpga}.

Perhaps the most popular and effective optimization for LSTMs is the use of reduced precision via linear quantization, where precision means the number of bits employed to encode inputs and weights. TPU~\cite{jouppi2017TPU} employs 8-bit weights and inputs for LSTM inference. Other proposals, such as Stripes~\cite{Stripes2016} and Bit Fusion~\cite{8416871}, support variable precision to further improve performance and energy efficiency for LSTM networks that can be computed with less than 8 bits. Despite the additional flexibility of these accelerators, the precision for each LSTM network is determined offline and it is fixed during inference. In other words, different LSTM networks can be evaluated at different precision, but a given LSTM is always computed at the same precision for all the inputs. In this work we propose a mechanism to dynamically select precision during inference of each individual LSTM to boost performance without any loss in accuracy.

In order to find an effective scheme to set the precision online, we analyzed the impact of the precision on the state of the LSTM cell. The cell state is the key component of an LSTM network as it stores information from previous inputs that will be used for future predictions. It consists of an array of N elements, where each element is computed by four neurons in different gates, i.e. fully-connected layers. Figure~\ref{f:cellstate_peak_no_fix} shows the evolution of one element in the cell state in a speech recognition network~\cite{miao2015eesen}, at three different levels of precision (32-bit floating point, 8-bit integer and 4-bit integer). As it can be seen, 8-bit quantization closely tracks the behavior of the 32-bit fp version, resulting in the same accuracy. However, 4-bit quantization introduces significant errors in some time steps, resulting in noticeable accuracy loss. Previous schemes would conclude that this LSTM network layer cannot be evaluated using 4 bits. However, a more detailed look at Figure~\ref{f:cellstate_peak_no_fix} reveals that the 4-bit version is actually able to mimic the behavior of the 32-bit version for a large percentage of time steps. More specifically, for phases where the cell state is stable the 4-bit version is quite accurate, whereas for phases where the cell state changes rapidly, i.e. peaks/valleys, it tends to exhibit a larger error. A more extensive analysis by using different LSTM networks and their respective training datasets shows that this behavior is actually quite prevalent: for stable phases the 4-bit version introduces a very small error of 19.6\%, whereas for peaks/valleys it introduces a large error of 78\% on average. For the sake of brevity, we will use the term peak to refer to both peaks and valleys.

Based on this observation, we propose a scheme that dynamically selects the appropriate precision by monitoring the state of the LSTM cell. Our system keeps track of the values of each element in the cell state in recent time steps. If the value is stable, the lowest precision supported by the hardware is selected to evaluate the next time step. Otherwise, higher precision is used (8 bits) to avoid significant errors during peaks. In our set of LSTM networks, this simple scheme allows us to use the lowest precision for more than 66\% of the time without any loss in accuracy. Note that our scheme not only dynamically changes the precision of a given cell element but it also uses different precision for different elements of a given cell.

\begin{figure}[t!]
	\centering
	\includegraphics[width=3.375in]{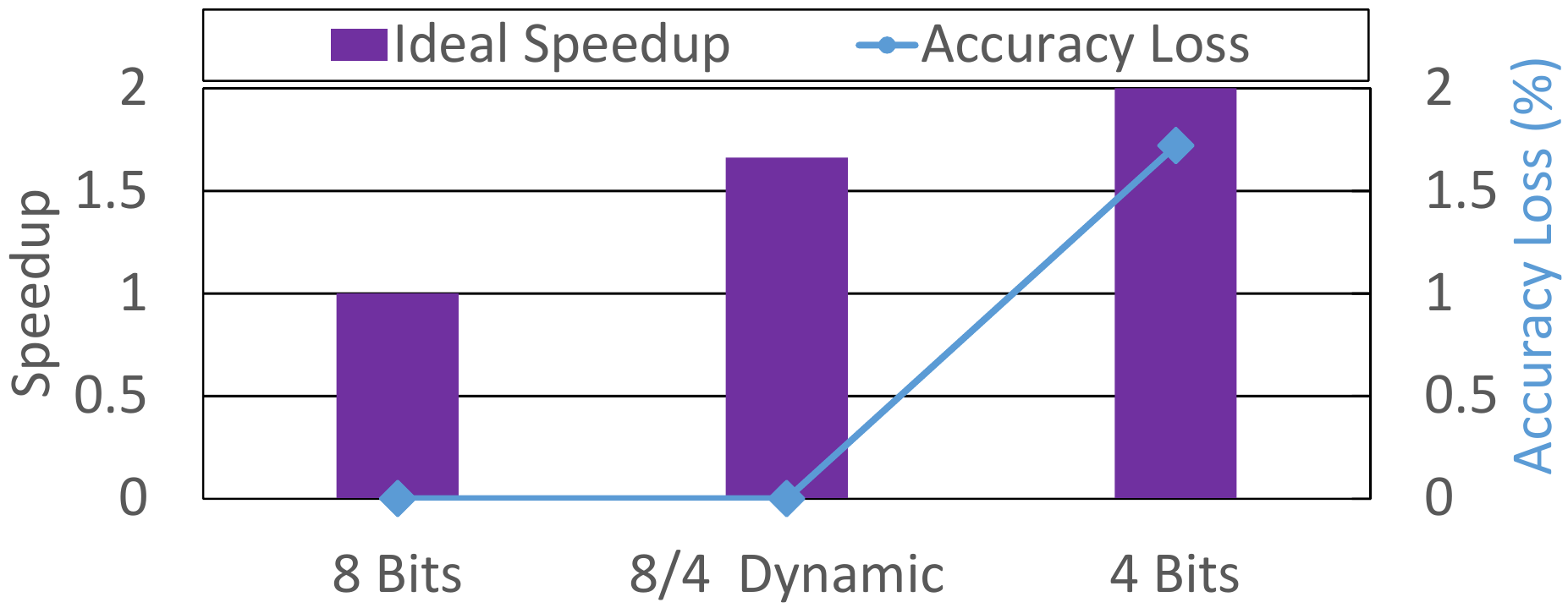}
	\caption{Speedup and accuracy loss for a speech recognition LSTM network~\cite{miao2015eesen} for different configurations. 8 Bits and 4 Bits are configurations that fix the precision to 8 and 4 bits respectively for all the time steps. 8/4 Dynamic is our scheme for dynamically selecting the precision online, that employs 4 bits for stable regions of the cell state and 8 bits for peaks. As it can be seen, our dynamic scheme outperforms the 8 Bit version without any loss in accuracy.}
	\label{f:ideal_speedup}
\end{figure}

We implement our scheme on top of E-PUR~\cite{fsilfa2018}, a recent accelerator highly optimized for LSTM inference. In order to support variable  
precision, the parallel dot product units in E-PUR are changed to Serial Inner Product (SIP) units. Then, we implement our dynamic precision selection scheme to decide the precision level for each element of the cell state on each time step. Our scheme provides 1.56x speedup and 23\% energy savings on average over the baseline, without affecting the accuracy. The extra hardware required for our technique introduces a small area overhead of 8.8\%.

Figure~\ref{f:ideal_speedup} illustrates the benefits of dynamic precision selection. By using 4-bit quantization, execution time would be reduced to approximately 50\% of the 8-bit version if it was used in all the time steps. However, it would introduce a significant degradation in accuracy. Our scheme, illustrated in the second bar, restricts the use of 4-bit quantization to stable phases of the cell state, which represent more than 66\% of the time, whereas peaks are evaluated using 8 bits. By doing so, we leverage 4-bit quantization for a large percentage of the execution, while avoiding any accuracy loss.

The main focus of this paper is high-performance and energy-efficient LSTM inference. We claim the following contributions:
\begin{itemize}
	\item We analyze the behavior of the LSTM cell state for a set of four popular LSTM networks. We conclude that for time intervals where an element of the cell state is stable, it can be evaluated with lower precision without any impact on accuracy, whereas peaks require higher precision to prevent accuracy loss.
	\item We propose a novel mechanism that uses the cell state in LSTM cells to dynamically select the appropriate precision for each time step and each cell element. Our scheme selects the lowest precision for more than 66\% of the time.
	\item We implement our technique on top of E-PUR, a state-of-the-art accelerator for LSTM inference. Our system improves performance by 1.56x and energy consumption by 23\% without loosing accuracy, while introducing a very small area overhead of 8.8\%.

\end{itemize}

%% file: background.tex
\section{Background}\label{s:background}

\subsection{Recurrent Neural Networks}\label{s:rnn_networks}
A Recurrent Neural Network (RNN) is a state-of-the-art machine learning algorithm that is very successful in sequence-to-sequence problems such as machine translation and speech recognition. Unlike conventional feed forward Deep Neural Networks (DNNs), RNNs include loops that allow them to store information from previous executions. For this reason, they have more context information, which allows them to make better predictions. In addition, since they are executed recurrently for each element of the input sequence they are able to handle problems that have a variable input and/or output sequence length. These features made them very effective for sequence-to-sequence problems, for which they outperform DNNs~\cite{greff2016lstm, schuster1997bidirectional}. Capturing long term dependencies is a challenging task for basic RNNs (Vanilla RNNs)  because the information tends to dilute over time. To solve this issue, the Long Short Term Memory (LSTM)~\cite{hochreiter1997long} networks were proposed.

\subsubsection{LSTM Cell}\label{s:lstm_cell}

\begin{figure}[t!]
	\centering
	\includegraphics[width=3.375in]{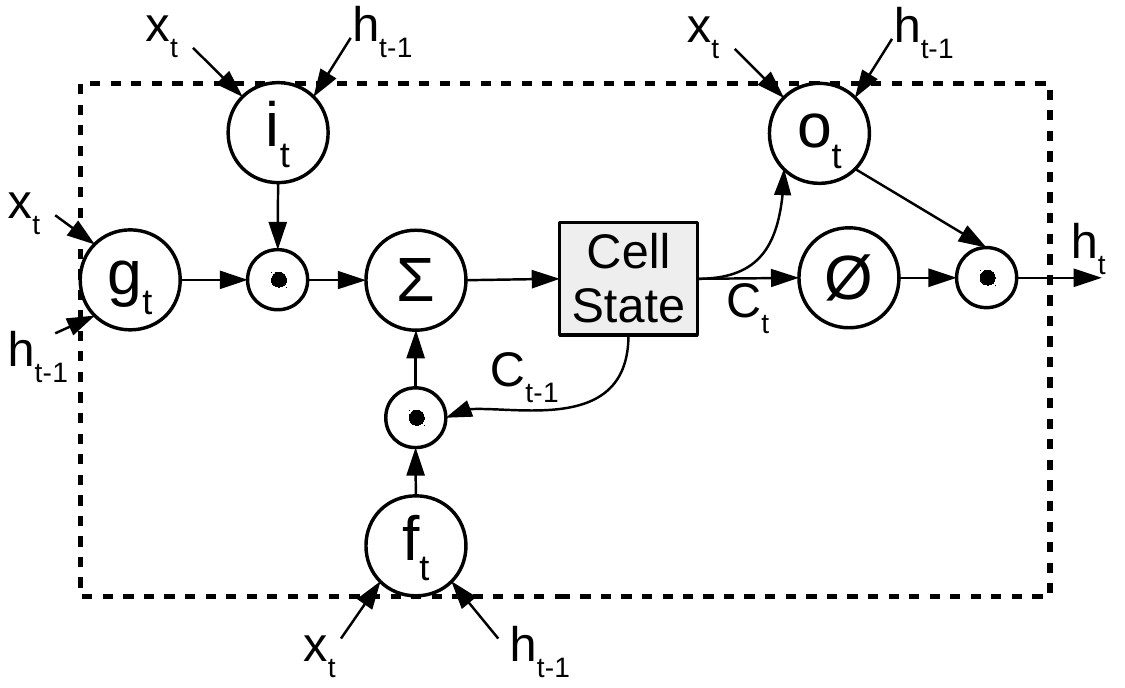}
	\caption{Structure of an LSTM cell. $\odot$ denotes an element-wise 
		multiplication of two vectors. $\phi$ denotes the hyperbolic
		tangent.}
	\label{f:lstm_cell}
\end{figure}

An LSTM network is composed of several LSTM cells stacked together to create a deep LSTM network. For 
each LSTM cell the main component is the cell state which stores the context information from
previous input elements of the sequence. In addition, LSTM cells use gates to modulate how the cell state is updated.
The main structure of an LSTM cell is shown in Figure~\ref{f:lstm_cell}. As it can be seen,
an LSTM cell is composed of four gates. Among these gates, the updater gate ($g_t$), whose computations are shown in Equation~\ref{e:updater_gate}, modulates the amount of input information that is being considered candidate to update the cell state ($c_t$). The input gate($i_t$), shown in 
Equation~\ref{e:input_gate}, controls what information will be added to the cell state. Shown in Equation~\ref{e:forget_gate} is the forget gate used to determine what information will be deleted from the current cell state ($c_{t-1}$). Finally, the output gate ($o_t$), shown in Equation~\ref{e:output_gate}, decides what information from the cell is outputted to create the cell output ($h_t$) for the current time step.

The computations performed by each gate on an LSTM cell are shown in Figure~\ref{f:lstm_equations}.
 Each gate has two types of connections: the forward and the recurrent connections. The forward connections operate on the input coming from a previous cell ($x_t$). On the other hand, the recurrent connections operate on the input coming from the cell output in the previous time step 
($h_{t-1}$).  The output of each gate is a vector and, for the sake of simplicity, we called the elements of this vector \textit{neurons}. In order to evaluate each of these neurons, an inner product between the weights on the forward connections and the vector $x_t$ is done. Then, the result is added to the inner product of the weights in the recurrent connections and the vector $h_{t-1}$. Finally, an activation function is applied which is normally a sigmoid or the hyperbolic tangent.

Most of the storage requirements of LSTM cells are due to the weight matrices and the output sequences. Regarding computations, most of the execution time is due to the evaluation of the matrix-vector multiplications of the various gates. 

In this work, we applied linear quantization to the weight matrices and the input vectors $x_t$ and $h_{t-1}$, as it is normally done to reduce storage and computing requirements, while the activation functions are evaluated in FP32.

\begin{figure}[t!]
	\centering
	\begin{align}
	i_t = \sigma(W_{ix} x_t + W_{ih} h_{t-1}  + b_i)
	\label{e:input_gate}
	\end{align}
	\begin{align}
	f_t = \sigma(W_{fx} x_t + W_{fh} h_{t-1}  + b_f)
	\label{e:forget_gate}
	\end{align}
	\begin{align}
	g_t = \phi(W_{gx} x_t + W_{gh} h_{t-1} + b_g)
	\label{e:updater_gate}
	\end{align}
	\begin{align}
	c_t = f_t \odot c_{t-1} + i_t \odot g_t
	\label{e:cell_state}
	\end{align}
	\begin{align}
	o_t = \sigma(W_{ox} x_t + W_{oh} h_{t-1}  + b_o)
	\label{e:output_gate}
	\end{align}
	\begin{align}
	h_t = o_t \odot \phi(c_t)
	\label{e:cell_output}
	\end{align}
	\caption{ LSTM cell computations. $\odot$, $\phi$, and $\sigma$ denote element-wise multiplication,
		hyperbolic tangent and sigmoid function respectively.}
	\label{f:lstm_equations}
\end{figure}

\subsection{Linear Quantization}\label{s:lineal_quantization}

Linear quantization is a commonly used technique to reduce memory footprint and computational
cost. The main idea consists of approximating a full precision value ($y$) to a value $y_q$ that is computed using an integer index and a quantization step ($q$) as shown in the following equations: 

\begin{equation}
q = \frac{\alpha}{2^{n-1}}
\label{e:quantization_step}
\end{equation}
\begin{equation}
i_k = round(y/q)
\label{e:quantization_index}
\end{equation}
\begin{equation}
y_q = q*i
\label{e:quantization_aprox}
\end{equation}

where $n$ is the bit width of the integer index $i_k$, f.e. 8 bits, and $\alpha$ is the
maximum absolute value of $y$. 

In the case of an LSTM gate, once the weights and inputs have been quantized, the computations for the output of a neuron ($z_k$) are done using integer arithmetic and the result is converted back to floating point as shown in Equation~\ref{e:quantized_inner_product} and Equation~\ref{e:quantized_inner_product_fp}:

\begin{equation}
z_i = \sum{w_i*x_i}
\label{e:quantized_inner_product}
\end{equation}

\begin{equation}
z_k =z_i*q_x*q_w
\label{e:quantized_inner_product_fp}
\end{equation}

where $w_i$ and $x_i$ are the quantized index for each element of the weight and input vector. Moreover, $q_x$ and $q_w$ are the quantization steps for the weight and inputs respectively. Multiplications are performed using 8 bits multipliers while summations and accumulations are normally performed with a higher number of bits (e.g., 24 bits).

\subsection{Serial Inner Products}\label{s:sip_products}
Serial Inner Products units (SIPs) have been previously proposed and used~\cite{Stripes2016} as a mechanism to exploit precision variability in different layers of a neural network. In a SIP unit, an inner product is computed by serially feeding the bits of one of the operands while the bits of the other are feed in parallel. On a cycle, a SIP unit performs the element-wise multiplications between a vector with \textit{1-bit} elements and a vector with \textit{n-bit} bits elements. These multiplications are typically done using AND and SHIFT operations. Then, the summations are performed by accumulating the partial products computed on each cycle. Note that, since only one bit from each element of one of the operands is multiplied on a given cycle, decreasing the bit width of the elements in the vector that is being fed serially will result in a linear increase for the SIP performance (i.e., reducing the bit width from 8 to 4 will result in 2x speedup).

Because bits are fed serially, for \textit{n-bit} numbers a SIP unit will take \textit{n} cycles to perform an inner product, hence effectively decreasing the overall bandwidth with respect to a system with parallel multipliers by a factor of \textit{n} (assuming that n-bit integer multiplication can be performed in one cycle with a conventional multiplier). A typical solution to address this issue is to increase the number of SIP units in a way that after \textit{n} cycles the system with parallel multipliers and the system with SIP units perform the same number of inner products, i.e. the number of serial products done in parallel in the SIP unit is increased to match the throughput of the parallel multiplier. To maintain each SIP unit busy, this approach requires that the problem has a large number of parallel computations. Since this is the case for LSTM networks, whose matrices-vectors being multiplied tend to be large and provide enough computations to keep the pipeline full, this is the approach adopted in this work to support variable precision.

%% file: techniques.tex
\section{Dynamic Precision Selection}\label{s:tecniques}

\begin{figure}[t!]
	\centering
	\includegraphics[width=3.375in ]{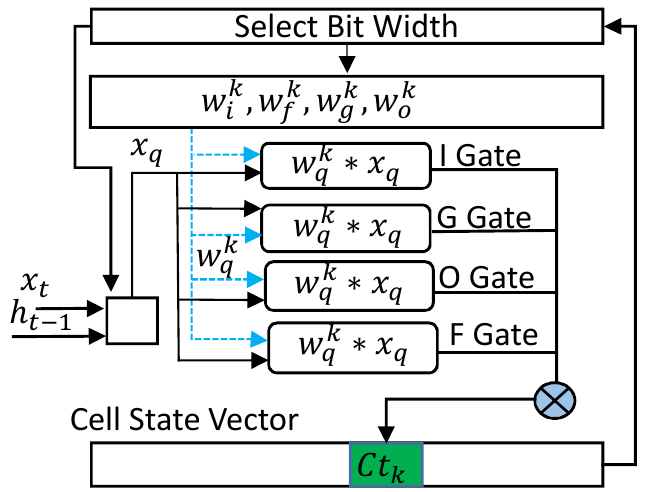}
	\caption{Relationship among neurons in the four gates and elements in the LSTM cell state. The value $Ct_k$ of element $n_k$ is computed based on the outputs of the $k^{th}$ neuron in each of the four gates. The precision used to evaluate those neurons is based on the evolution of the element $n_k$.}
	\label{f:neuron_to_cell_state_mapping}
\end{figure}

\begin{figure}[t!]
	\centering
	\begin{align}
		r = maxC_i -minC_i
		\label{e:peak_range}
	\end{align}
	\begin{align}
		upperLimit = r+r*\beta
		\label{e:peak_upper_limit}
	\end{align}
	\begin{align}
		lowerLimit = r-r*\beta
		\label{e:peak_lower_limit}
	\end{align}
	\begin{align}
     	isInPeak = lowerLimit \leq c_i \leq upperLimit 
	\label{e:peak_is_inpeak}
	\end{align}
		\caption{ Positive and negative peak definition for the cell state of a given neuron. 
	     }
	\label{f:peak_definitions}
\end{figure}

In this section we describe the proposed scheme to dynamically adjust the number of bits used
to encode and operate the input and weight vectors on LSTM networks, with the goal of increasing performance 
and reducing the energy consumption of the system. First, we discuss the main
performance and energy bottlenecks on state-of-the-art hardware accelerators 
for LSTM inference. Next, we present the key idea for our 
precision selection scheme. Finally, we describe the hardware 
implementation of our technique.

\begin{figure*}[t!]
	\centering
	\includegraphics[width=6.5in]{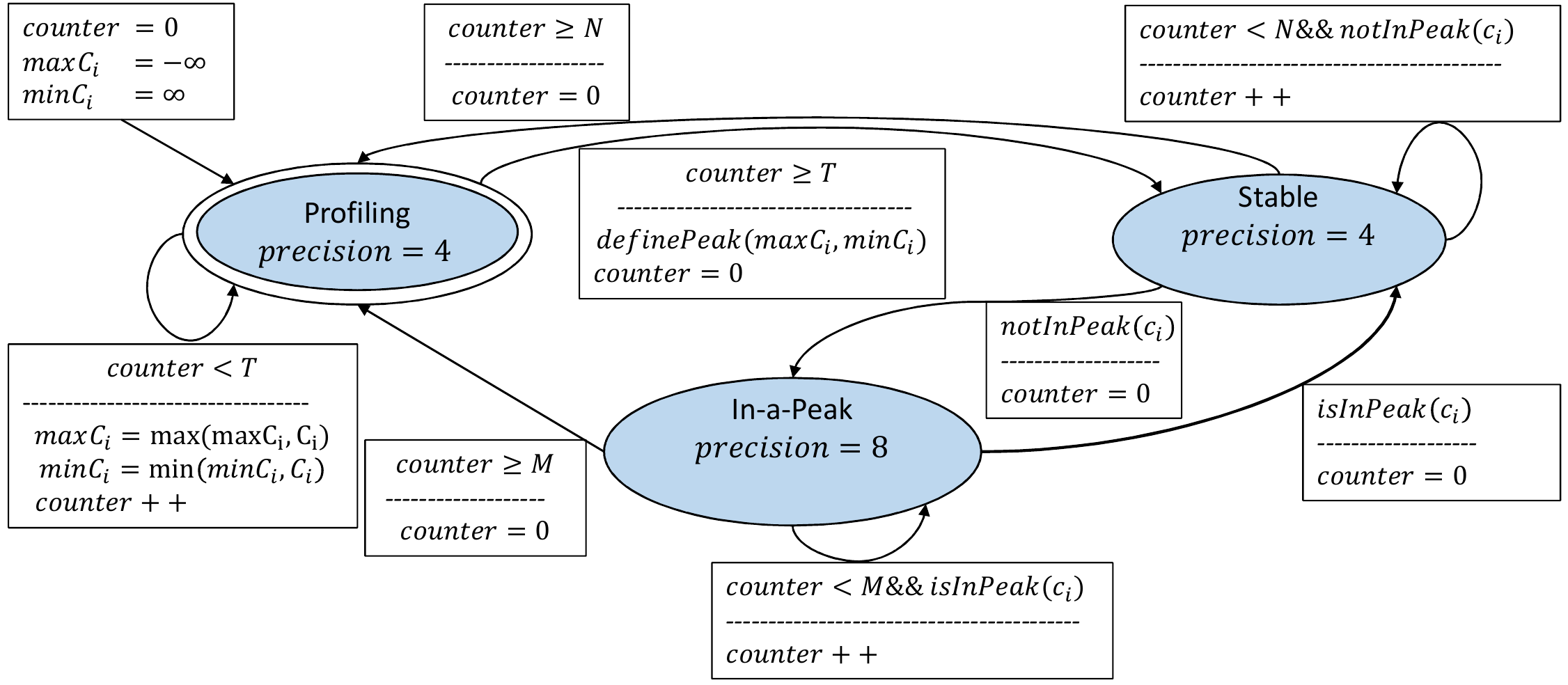}
	\caption{State machine employed to dynamically select the precision for each element in the LSTM cell state.}
	\label{f:state_machine}
\end{figure*}

\subsection{Motivation}\label{s:precison_selection_motivation}
LSTM cells are composed of four gates, each one with two matrices containing the weights for the forward and the recurrent connections respectively. Since these weight matrices tend to be quite large, most of the energy consumed by state-of-the-art hardware accelerators for LSTM inference is due to the static and dynamic energy consumed by the memories employed to store the weights and intermediate results. Not surprisingly, the energy consumption of these on-chip memories accounts for up to 80\% of the total energy in state-of-art solutions for LSTM~\cite{fsilfa2018}.  

An effective way to decrease memory footprint and thus static and dynamic energy without affecting accuracy is using Linear Quantization (see Section~\ref{s:lineal_quantization}). Normally, a static profiling of the network is done in order to determine the minimum precision that can be used to quantize an LSTM model without losing accuracy. A common approach is to set a fix bit precision (i.e. 8 bits) for the whole network. However, while this solution covers the worst case, it ignores cases where a lower precision could be employed for a subset of computations without losing accuracy. Figure~\ref{f:ideal_speedup} shows the accuracy loss of an LSTM network for speech recognition~\cite{miao2015eesen} when using 8 bits, 4 bits and a mix of both. In the case of the mixed precision, for each individual neuron (e.g., inputs and weights), we dynamically set the precision to 8 or 4 bits as described later in Section~\ref{s:dynamic_presicion_overview}. As it can be seen in Figure~\ref{f:ideal_speedup}, using a precision of 8 bits (i.e., assuming worst case precision for the whole network) results in no accuracy loss. On the contrary, using a precision of 4 bits results in 2\% of accuracy loss (which is an important loss for speech recognition). However, it can be observed that more that 50\% of the computations can be evaluated using 4 bits while the rest are computed using 8 bits without losing accuracy. Note that previous work have reported and exploited variability in the precision requirements for DNN computations~\cite{Stripes2016, Judd2015}. However, they exploited precision variability across layers whereas in this work we focus in precision variability among neurons and time steps of execution, which is a much finer-grain variability. In other words, prior work supports different precision for different DNNs, but the precision for each DNN and layer is determined offline and kept constant during inference. Our proposal is different as we dynamically select the precision for each neuron and time step.

A main challenge to dynamically change the precision is deciding when to use a high or low 
precision. In this work, we propose to use the state of the LSTM cell as an indicator of the required precision. More specifically, we propose to use high precision (i.e. 8 bits) for evaluations performed when a cell state element is in a peak and low precision (i.e. 4 bits) for the rest. 

We base this proposal on the key observation that when an element of the cell state is stable (i.e. is changing slowly) the absolute difference between the cell state evaluated in full precision ($c_{t}^fp$) and the cell state computed in low precision ($c_{t}^l$) tends to be small. On the contrary, when the cell state is changing fast (i.e. in a peak) the difference between $c_{t}^l$ and $c_{t}^fp$ tends to be very large, which introduces a significant error that results in accuracy loss. This behavior is illustrated in Figure~\ref{f:cellstate_peak_no_fix}, that shows the cell state for a given neuron evaluated using different precision for multiple consecutive time steps of execution. As it can be seen, for several peaks the error in the cell state evaluated using low precision (4 bits) tends to be quite large: more than 70\% error when compared with the cell state computed with full precision in FP32 format. However, outside the peaks, i.e. in stable phases of the cell state, the error tends to be small. More specifically,
after an exhaustive profiling of different LSTM networks, we found that on average the relative error in the peaks is 78\% whereas the error in stable regions is 19.6\%. Therefore, since introducing a larger error into the cell state results in a larger accuracy loss, we propose to evaluate the peaks using high precision while performing the computations outside the peaks using a lower precision.

In summary, we design a scheme that monitors the evolution of the cell state at run-time for each element and selects high precision during peaks and low precision for stable phases. For this work, we use 8 bits for the high precision since it provides zero accuracy loss for all tested LSTM networks. On the other hand, we use 4 bits for the lower precision, which would have a significant loss in accuracy if it was used for all the time steps. In the following sections we detail this scheme and describe its hardware implementation on top of a state-of-the-art accelerator.

\newpage

\subsection{Overview}\label{s:dynamic_presicion_overview}

The main idea of our proposal is to set the precision at each time step of execution for the input vectors $x_t$ and $h_{t-1}$ and their corresponding weights for each single element of the cell state individually. For a given LSTM cell, the $k^{th}$ element of the cell state vector is computed using a combination of the output value of the $k^{th}$ neuron on each gate. We refer to these four neurons simply as element $n_k$ of the LSTM cell and set the precision for the four of them in tandem, since all of them are associated with the same element of the cell state. This relationship is shown in Figure ~\ref{f:neuron_to_cell_state_mapping}.

To determine when each element $n_k$ of the cell state is on a peak, we employ the state machine depicted in Figure~\ref{f:state_machine}. In order to track the evolution of the value of the cell state ($c_k$) of a given element $n_k$ we divide the process into three phases.
First, the system starts in a \textit{profiling state} that samples $c_k$ for a certain number of time steps. This profiling is done in order to determine the peak characteristics of $c_k$. Then, we have the \textit{stable state} that indicates that $c_k$ has had a stable value for the previously evaluated time steps. Finally, the \textit{in-a-peak state} tracks when $c_k$ is in a peak.

As shown in Figure~\ref{f:state_machine}, the \textit{profiling state} is performed for $T$ time steps. In each profiling step we keep track of the maximum and minimum value of cell state. Note that the profiling is done using low precision (4 bits) because we assume that while profiling the cell state is inside a stable region. Finally, after $T$ time steps, we use the maximum and minimum value of $c_k$ to set the limit values that define when a peak begins or ends, and then we move to the \textit{stable state}.

\begin{figure}[t!]
	\centering
	\includegraphics[width=3.375in]{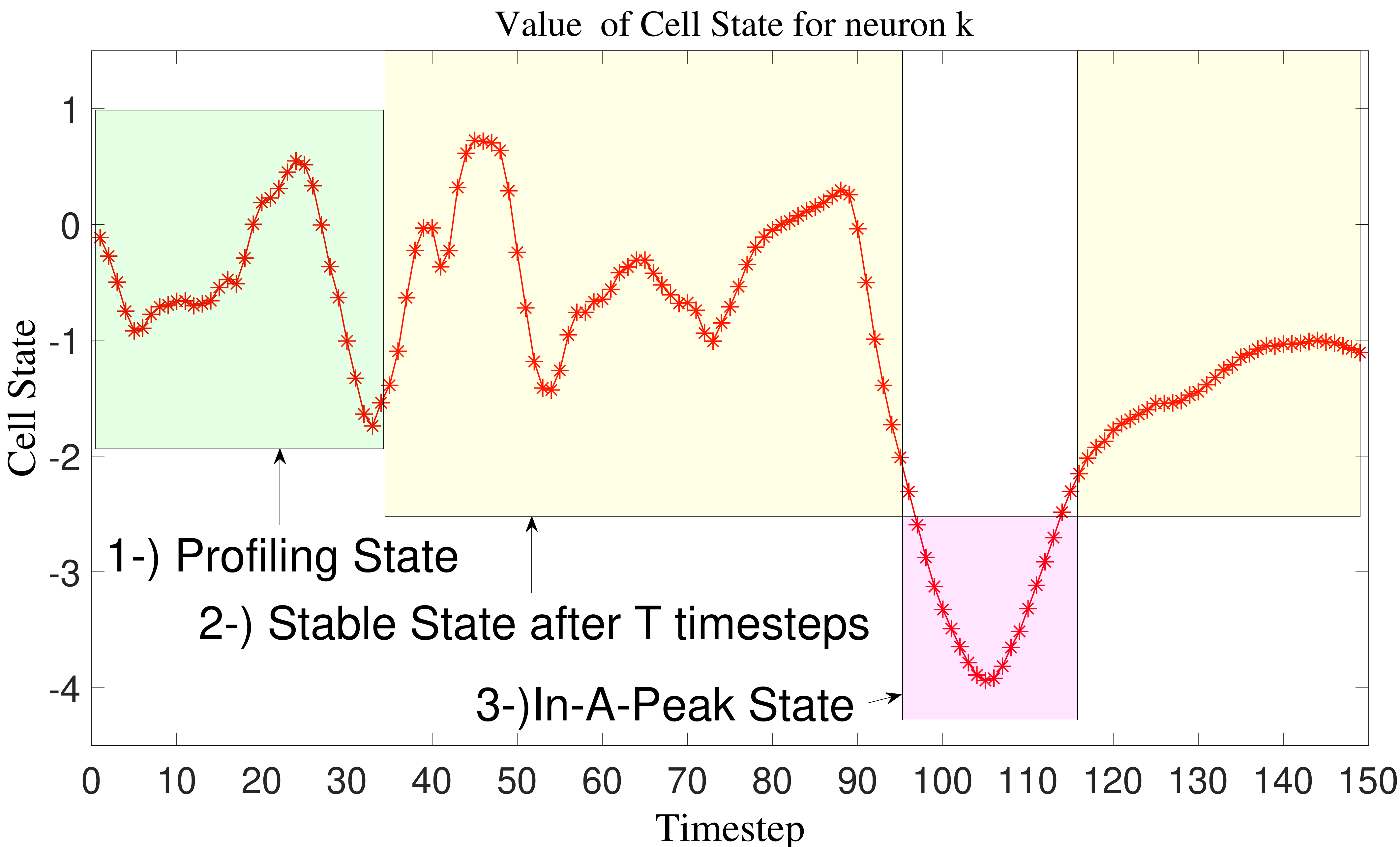}
	\caption{Evolution of the LSTM cell state for a given neuron. At each time step its stability is tracked in order to decide the precision for the next time step. }
	\label{f:example_state_machine}
\end{figure}

The system remains in the \textit{stable state} until a peak is detected. A peak is found using the values $minC_i$ and $maxC_i$, obtained previously in the profiling state, as shown in Figure~\ref{f:peak_definitions}. In order to determine that the value in the cell state has entered a peak, we require that it exceeds the $minC_i$ and $maxC_i$ found during the profiling stage by a given margin to increase the confidence of the detection. To this end, we use the parameter $\beta$ in Equation~\ref{e:peak_upper_limit} and Equation~\ref{e:peak_lower_limit} to establish the upper and lower thresholds. If the $c_k$ value in the cell state exceeds one of these thresholds, a peak is detected and the system transitions to the \textit{in-peak-state} to use high precision. It remains in this state until we detect that $c_k$ is no longer in a peak using Equation~\ref{e:peak_is_inpeak}. In case that the end of the peak is detected, we move to the \textit{stable state} to
switch back to low precision, as the value of the cell state has entered a stable phase.

If the system stays in a peak for a large number of time steps (e.g., $M$ in Figure~\ref{f:state_machine}), the profiling stage is triggered again. Note that this profiling is required since the $c_k$ may become stable at a value that is outside the thresholds of the original profiling and, in this case, our scheme would stay indefinitely in high precision state if profiling is not repeated to set new upper and lower thresholds. In other words, the information of the initial profiling may become outdated, since the range of values of the cell state may shift over time. On the other hand, the system may be stuck in the \textit{stable state} in case the range of the values of the cell state become narrower over time, as they will never exceed the minimum and maximum thresholds set in the initial profiling. To prevent this issue, we force a profiling stage when the system stays in the \textit{stable state} for more than $N$ time steps. By doing this, we take into consideration recent values of the cell state and more adequate thresholds are set.

\begin{figure}[t!]
	\centering
	\includegraphics[width=3.375in]{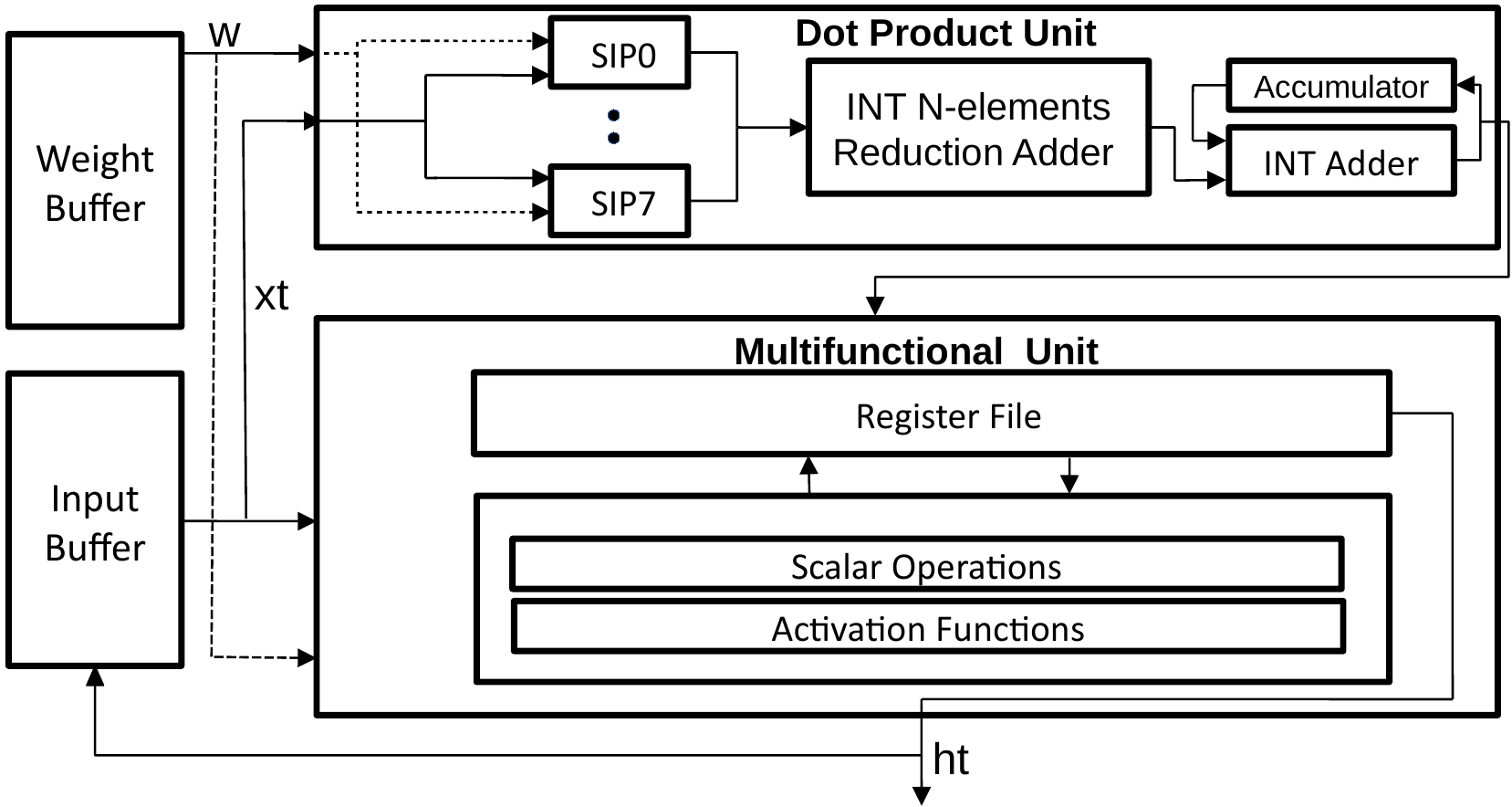}
	\caption{Compute Unit (CU). SIP units are included to support variable precision and they are replicated in order to match the throughput of conventional multipliers.}
	\label{f:computational_unit}
\end{figure}

 Our overall scheme for dynamic precision selection is summarized in Figure~\ref{f:example_state_machine}.
Considering an input sequence with elements $x_0$ to $x_{n-1}$, for a given cell state element $n_k$, the scheme works as follows. First, the value $c_k$ of the element $n_k$ is computed using low precision and the \textit{profiling state} is executed for $T$ time steps, marked as $1)$ in Figure~\ref{f:example_state_machine}.
Then, after the profiling stage, the system moves to the \textit{stable state} and performs all computations associated with $n_k$ using low precision. Then, at timestep $p$, we detect that $c_k$ is lower than its previously profiled lower threshold and, thus, the system changes to the \textit{in-a-peak state}, as seen in Figure~\ref{f:example_state_machine}. Next, it stays in this state until the value of $c_k$ comes back to its previously profiled range and, then, it switches back to \textit{stable state}, where it waits for the occurrence of another peak or the triggering of another profiling stage. Note that this process is performed for each element in the LSTM cell state individually and, hence, our system may select different precision for different neurons in the same time step.

\subsection{Hardware Implementation}\label{s:hardware_implementation}

We implement our dynamic precision selection scheme on top of E-PUR~\cite{fsilfa2018}, a state-of-the-art low power accelerator for LSTM networks~\cite{fsilfa2018}. E-PUR consists of several computational units that evaluate the four different gates in an LSTM cell. Furthermore, it includes on-chip storage for weights and intermediate results. In order to support variable precision, we replace the parallel multipliers by the SIP units. In the next subsections we describe the overall architecture of E-PUR and explain how it can be extended to implement our scheme for dynamic precision selection.

\subsubsection{Supporting Variable Precision}\label{s:hardware_baseline}

We extended E-PUR to use SIP units instead of parallel multipliers to compute inner products. Figure~\ref{f:computational_unit} shows the structure of a computational unit (CU) which is tailored to the evaluation of a gate in an LSTM cell. A CU is composed of a dot product unit (DPU), a Multi-Functional Unit (MU) and several buffers to store the weights and inputs. DPUs are used to compute the matrix-vector multiplications in the four gates. In the original E-PUR implementation, an inner product of $K$ numbers is done on each cycle and multiplications are performed using 8 bits. Therefore, to maintain the same throughput, 8 SIP units are included per DPU. 

\begin{figure}[t!]
	\centering
	\includegraphics[width=3.375in]{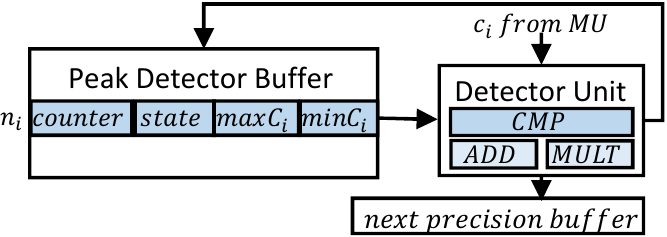}
	\caption{Structure of the Peak Detector Unit (PDU). }
	\label{f:peak_detector_unit}
\end{figure}

\begin{figure}[t!]
	\centering
	\includegraphics[width=3.375in]{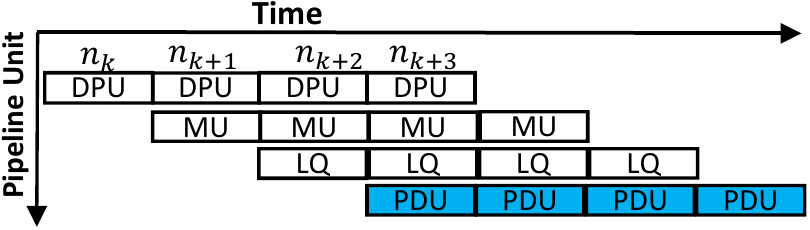}
	\caption{Overlapping of computations in the accelerator. 
	}
	\label{f:exectuion_pipeline}
\end{figure}

MUs are used to compute activation functions and scalar operations using floating point numbers. In addition, they are used to do the quantization of the cell output ($h_{t-1}$) and to convert the DPU output to floating point. Note that weights are quantized offline.

\begin{figure}[t!]
	\centering
	\includegraphics[width=3.375in]{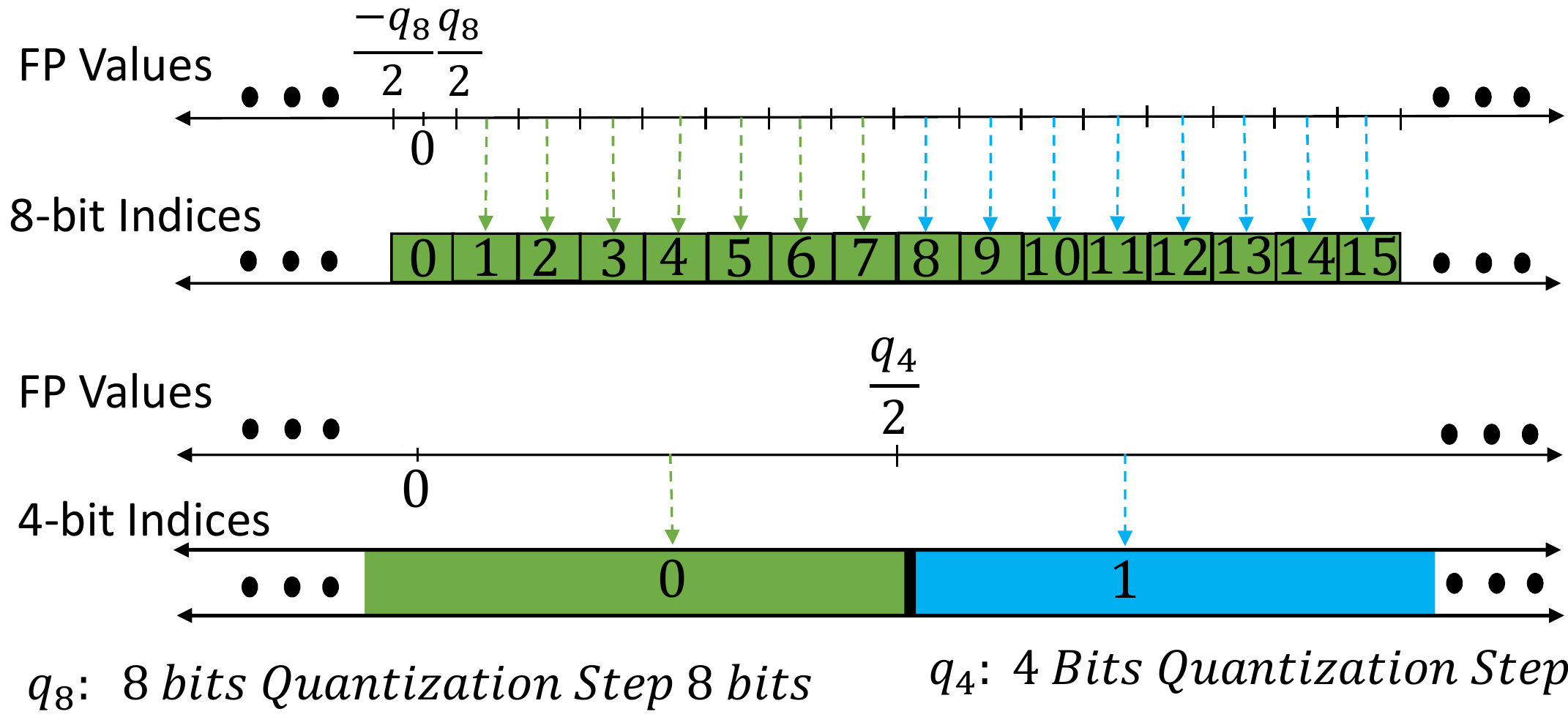}
	\caption{Linear Quantization of floating point values for 8 and 4 bits. For some cases, using the most significant nibble of an 8-bit index to represent the corresponding 4-bit index yields to an incorrect mapping. }
	\label{f:encoding_byte}
\end{figure}

In E-PUR+SIP, the gates in an LSTM cell are evaluated in parallel whereas neurons are computed in a sequential manner for the inputs $x_t$ and $h_{t-1}$. For a given element, i.e. $n_k$, the following steps are followed in order to compute its output $y_t$. First, the input ($x_t$) and its weight vector are split into $K$ sub-vectors of size $N$.
Then, on each CU, 8 sub-vectors of weights with $N$ elements are fetched from the weight buffer and dispatched to each SIP unit. Furthermore, an 8*$N$-bit vector ($v_0$), corresponding to the most significant bit of each element in $x_t$, is fetched from the input buffer and dispatched to each SIP unit to perform the multiplication of $v_0$ with the corresponding weights. After this, each SIP accumulates its output and the next vector of bits ($v_1$), which corresponds to bit 6 of each element in $x_t$, is fetched from the input buffer and dispatched to each SIP, where they are multiplied by their corresponding weights. Next, the results of each SIP are added together with the previously accumulated output. This process is repeated until all the bits in $x_t$ are multiplied and added together. Finally, the accumulated values on each SIP unit are added together and the the process is repeated for the remaining sub-vectors.

\begin{table*}[t!]
	\caption{LSTM Networks used for the experiments. Low Precision Usage column shows the percentage of the evaluations performed at low precision.}
	\label{t:lstm_networks}
	\centering
	\begin{tabular}{ccccccc}
		\cellcolor[gray]{0.9}\small\textbf{Network}&\cellcolor[gray]{0.9}\small\textbf{App Domain}&\cellcolor[gray]{0.9}\small\textbf{Layers}&\cellcolor[gray]{0.9}\small\textbf{Cell Size}&\cellcolor[gray]{0.9}\small\textbf{Base Accuracy}&\cellcolor[gray]{0.9}\small\textbf{ Low Precision Usage }&\cellcolor[gray]{0.9}\small\textbf{ Dataset }  \\
		\small IMDB Sentiment~\cite{daiL15a}&\small Sentiment Classification&\small 1&\small 128&\small86.5\%&\small 100\% &\small IMDB dataset \\
		\cellcolor[gray]{0.9}\small SHOW TELL~\cite{vinyalsTBE16}&\cellcolor[gray]{0.9}\small Image Description&\cellcolor[gray]{0.9}\small 3&\cellcolor[gray]{0.9}\small 512&\cellcolor[gray]{0.9}\small 32.2 Bleu&\cellcolor[gray]{0.9}\small 52\%&\cellcolor[gray]{0.9}\small  MSCOCO\\
		
		\small EESEN~\cite{miao2015eesen}&\small Speech Recognition&\small 10&\small 320&\small 23.8     WER&\small 70\%&\small Tedlium V1\\
		\cellcolor[gray]{0.9}\small MNMT~\cite{wu2016google}&\cellcolor[gray]{0.9}\small Machine Translation&\cellcolor[gray]{0.9}\small 8&\cellcolor[gray]{0.9}\small 1024&\cellcolor[gray]{0.9}\small26.0 Bleu&\cellcolor[gray]{0.9}\small 44\%&\cellcolor[gray]{0.9}\small WMT'15 En $\rightarrow$ Ge \\
	\end{tabular}
\end{table*}

Once the output value $y_t$ is computed, it is sent to the MU, where the value is converted to floating point and the activation function is computed. After each gate is evaluated, the cell state is computed and stored in the input buffer by the output gate. In addition, the output value $h_t$ is computed and quantized. Finally, the MU stores the final result in the on-chip memory for intermediate results. Note that the operations to compute the dot product, the activation function and to quantize the result are overlapped, as seen in Figure~\ref{f:exectuion_pipeline}. Hence, once the DPU sends a result to the MU, it will continue with next neuron. Similarly, once an activation function is computed, we proceed with the computation of the next activation while applying the quantization steps to the previously computed activation. These steps are repeated until all the neurons in the LSTM cell are evaluated.

\newpage
\subsubsection{Dynamic Precision Selection }\label{s:dynamic_precision_support}

In order to set the precision for each neuron at each time step, we extend EPUR with a Peak Detector Unit (PDU), as shown in Figure~\ref{f:peak_detector_unit}. This unit tracks the evolution of each element in the LSTM cell state. As seen in Figure~\ref{f:peak_detector_unit}, the PDU is composed of a buffer that stores the information needed by the state machine shown in Figure~\ref{f:state_machine}. Furthermore, it includes a detector unit that is employed to detect when the tracked cell state is inside or outside a peak according to Equation~\ref{e:peak_is_inpeak}. The PDU updates the state for a given element of the cell state after the MU computes its value and set the precision to be used in the next time step in the \textit{next precision buffer}. Note that the computations performed by the PDU are overlapped with the MU evaluations in a pipelined manner, as shown in Figure~\ref{f:exectuion_pipeline}.

One major challenge to dynamically set the precision is storing the quantized integer indices for the weight matrix and input vectors in an energy efficient manner. This challenge arises because at each time step, an index can be fetched in either low or high precision. Therefore, a mechanism that is able to store and fetch both indices in an energy efficient way is needed. One possible solution is that for a given floating point value the indices for high and low precision are store separately. The main drawback for this approach is that the memory footprint increases by 50\% and, hence, increasing significantly the energy consumption of the system.

A more energy efficient alternative is to use only one byte to store the high and low precision indices for a single weight. In this approach, for a given floating point value which is quantized in low and high precision, the most significant nibble of its high precision index is also used to store its low precision index. Therefore, the memory footprint of the baseline system is not increased. In addition, if the most and least significant nibbles of a given index are stored separately, only half of the memory accesses are needed to fetch the low precision indices, hence dramatically decreasing the dynamic energy consumption of the system. However, we found that this straightforward solution has a negative impact in the accuracy. In our experiments, the accuracy loss is larger than 50\% for our set of LSTM networks. 

Figure~\ref{f:encoding_byte} shows a mapping of some floating-point numbers to integer indices using 8 bits (top of the figure) and 4 bits (bottom of the figure). As it can be seen, using just the most significant nibble of the 8-bit index to obtain the 4-bit index is incorrect in some cases. As an example, a floating point value mapping to index $11$ in 8 bits is quantized as 1 when using 4 bits. However, using the most significant nibble of the 8-bit index would give an incorrect value of 0. A key observation is that values which are incorrectly mapped to its low precision counterpart always have a difference of one with the correct representation. Therefore, the correct index can be obtained by adding one to the most significant nibble of the corresponding high precision index. Note that we only show in Figure~\ref{f:encoding_byte} the first 16 indices. Negatives values are omitted for the sake of simplicity. However, the same issue would arise for the rest of indexes.

In this work, we use the most significant nibble of the high precision indexes to obtain its low precision counterparts. We include an extra bit (\textit{offset bit}) that is set to one for indexes that are mapped incorrectly. Furthermore, we include the extra hardware required to increment by one unit those indexes. Note that weights are static and, hence, their \textit{offset bits} can be set offline. Regarding the input elements, their \textit{offset bit} is set online using a small table where each entry indicates if the \textit{offset bit} for an index is set to one or zero. We account for this extra hardware in our experimental results.

In order to evaluate an LSTM cell, in addition to the steps described in Section~\ref{s:hardware_baseline}, we also perform some extra tasks to set the appropriate precision for each neuron. First, for a given element $n_i$, the precision to be used is read from the \textit{next precision buffer} in the PDU. Then, if high precision is chosen, we fetch a byte for each element in the weight vector for $n_i$. On the other hand, if low precision is selected, we only fetch the most significant nibble. In this case, we also read the \textit{offset bit} in order to know if we need to adjust it, increasing its value by one unit to obtain the 4-bit index. Once the values have been fetched and adjusted, we send them to each of the SIP units as outlined in Section~\ref{s:hardware_baseline}.
 
Regarding the input vector, we first fetch all the elements in the input vector in high precision. Next, if low precision is selected, we adjust the elements of the vector that have its \textit{offset bit} set to one and then we proceed to feed them serially to each SIP unit. Since the same input vector is used for all the neurons, we cache the input vector encoded in low precision after it is adjusted.

Finally, once we compute the cell state in the MU, it is sent to the PDU in order to determine the precision to be used on element $n_i$ in the next time step. In addition, the output value is sent to the quantization unit where it is quantized and we set the \textit{offset bit}. Note that these operations are overlaped as shown in Figure~\ref{f:exectuion_pipeline}, and their latency is hidden by the computations in the DPU.


%% file: methodology.tex
\begin{table}[t!]
	\caption{Hardware Configuration.}
	\label{t:epur_sip_params}
	\centering
	\begin{tabular}{cc}
		\hline
		\multicolumn{2}{c}{\textbf{E-PUR+SIP }}\\
		\cellcolor[gray]{0.9}\small\textbf{Parameter}&\cellcolor[gray]{0.9}\small\textbf{Value}\\
		\small Technology&\small 28 nm\\
		\cellcolor[gray]{0.9}\small Frequency&\cellcolor[gray]{0.9}\small 500 MHz\\
		\small Intermediate Memory&\small 6 MiB\\
		\cellcolor[gray]{0.9}\small Weight Buffer&\cellcolor[gray]{0.9}\small 2 MiB per CU\\
		\small Input Buffer&\small 8 KiB per CU\\
		\cellcolor[gray]{0.9}\small DPU Width&\cellcolor[gray]{0.9}\small 16 operations\\
		\small MU Operations&\small cycles: 2 (ADD), 4 (MUL), 5 (EXP) \\
		\cellcolor[gray]{0.9}\small MU Communication&\small \cellcolor[gray]{0.9} 2 cycles\\
		\small Peak Bandwidth&\small  30 GB/s \\
		\cellcolor[gray]{0.9}\small Peak Detector Buffer&\cellcolor[gray]{0.9}\small 8 KiB \\
		
		\hline
		\multicolumn{2}{c}{\cellcolor[gray]{0.9}\textbf{State Machine Configuration}}\\
		
		\small  M &\small 5\% of time steps\\
		\cellcolor[gray]{0.9}\small N &\cellcolor[gray]{0.9}\small 5\% of time steps\\
		\small $\beta$ &\small 0.1\\
		
		\hline
		
	\end{tabular}
\end{table}

\newpage
\section{Evaluation Methodology}\label{s:methodology}

To evaluate our proposal we employ a diverse and representative set of modern LSTM networks as shown in Table~\ref{t:lstm_networks}. We include four LSTM networks from popular real-world applications: speech recognition~\cite{miao2015eesen}, machine translation~\cite{wu2016google}, image description~\cite{vinyalsTBE16} and sentiment classification~\cite{daiL15a}. Our benchmarks largely differ in the number of internal layers and the dimensions of the cell size. For LSTM inference, we feed the LSTM networks with inputs from their respective test datasets, that include thousands of input sequences for each network. The length of the input sequences ranges from 20 time steps to a few thousands.
The LSTM networks were implemented in Tensorflow~\cite{tensorflow2016}. The original accuracy for each network, using 32-bit floating point arithmetic, is reported in Table~\ref{t:lstm_networks}.

In order to assess execution time and energy consumption, we employ a cycle-level simulator that models E-PUR~\cite{fsilfa2018} with the modifications presented in Section~\ref{s:hardware_implementation}. Furthermore, the dynamic precision selection scheme,  described in Section~\ref{s:dynamic_precision_support}, is implemented. The simulator provides the execution time and the activity factors of the different hardware components. Regarding the power model, the pipeline components are implemented in Verilog and synthesized using the Synopsys Design Compiler to obtain their energy consumption, area and delay. We use a typical process corner with voltage of 0.78V. In addition, we employ CACTI~\cite{muralimanohar2009cacti} to estimate the delay and energy consumption (static and dynamic) of the on-chip memories. Finally, to estimate timing and energy consumption of main memory, we use MICRON's memory model~\cite{Micron}. We model 4 GB of LPDDR4 DRAM. Regarding the frequency, we employ the delays reported by Synopsys Design Compiler and CACTI to set it: we specified a clock frequency that allows most hardware structures to operate at one clock cycle. The parameters of the accelerator used in our experiments are shown in Table~\ref{t:epur_sip_params}.

Since E-PUR is designed to accommodate LSTM models of different sizes, some of its on-chip storage and functional units might be oversized for some models. In this case, unused memory banks and functional units are power gated when not needed.

Regarding the Peak Detector Unit (PDU) and extra hardware required to support the dynamic precision, the configuration parameters are also shown in Table~\ref{t:epur_sip_params}. E-PUR includes DPU units that perform 16 8-bit multiplications per cycle and, hence, we replace them by 8 SIP units on each CU in order to match the throughput while supporting variable precision. In addition, the PDU requires a buffer of 8 KB to support the largest LSTM model.

The state machine for dynamic precision selection, shown in Figure~\ref{f:state_machine}, requires three different parameters: $M$ and $N$ control the maximum number of time steps that the system may remain in states \textit{In-a-Peak} and \textit{Stable} respectively before triggering the profiling, whereas $\beta$ is used to set the upper and lower thresholds to decide whether the value of the cell state is inside a peak. We performed a design space exploration for these parameters and found values that provide good results across the four LSTM networks used. Therefore, these parameters do not have to be manually tuned for each new LSTM network, as we empirically determined that the values shown in Table~\ref{t:epur_sip_params} provide excellent results for a wide variety of networks. Furthermore, to prove that these values generalize well for new unseen inputs, we performed the design space exploration by using the training datasets, whereas the evaluation of the technique is performed by using the test datasets.

\begin{figure}[t!]
	\centering
	\includegraphics[width=3.375in]{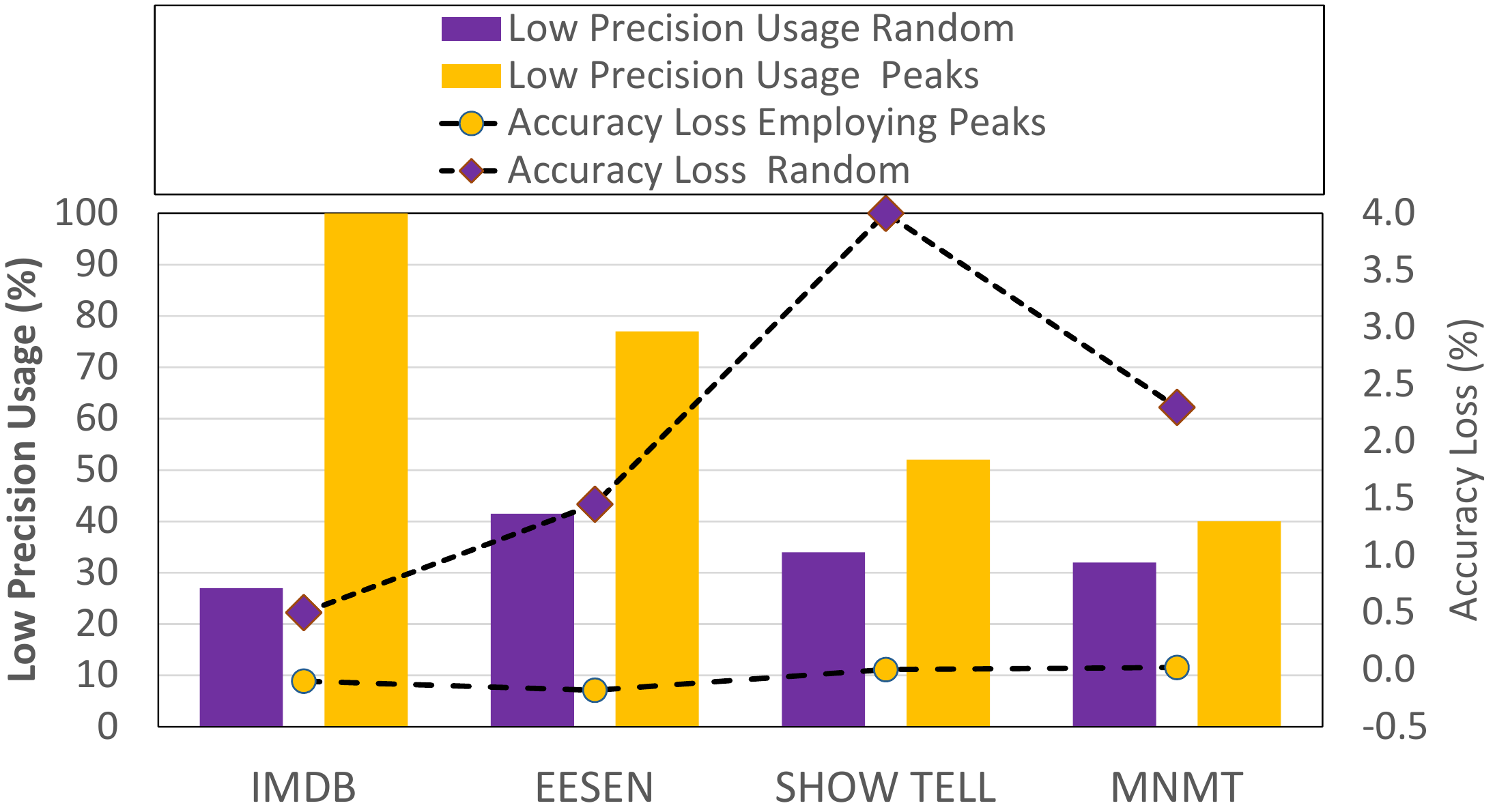}
	\caption{Comparison between our dynamic precision selection scheme (``Peaks'') and a system than randomly chooses the evaluations done at low precision (``Random''). The random scheme produces a significant degradation in accuracy, even if the percentage of evaluations in low precision is kept relatively low. Our scheme achieves much higher coverage without any accuracy loss.}
	\label{f:random_accuracy_loss}
\end{figure}

\begin{figure}[t!]
	\centering
	\includegraphics[width=3.375in]{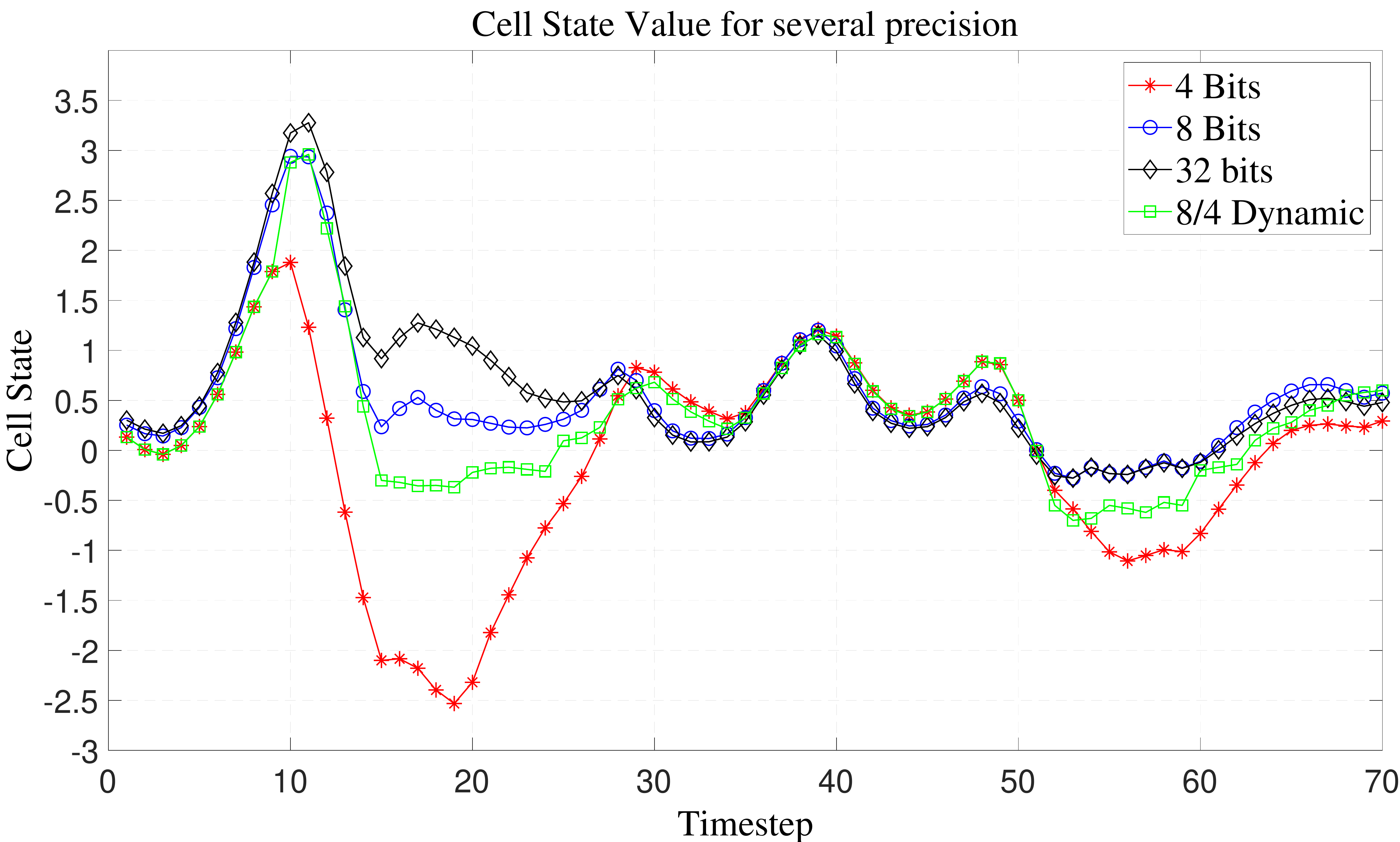}
	\caption{Evolution of one element in the cell state of a speech recognition LSTM network~\cite{wu2016google}. By using 8-bit precision in the peaks and 4-bit precision in the stable regions, we are able to track more accurately the full precision evaluation of the cell state.}
	\label{f:cell_state_peak_fix}
\end{figure}

%% file: results.tex
\section{Experimental Results}\label{s:results}

\begin{figure}[t!]
	\centering
	\includegraphics[width=3.375in]{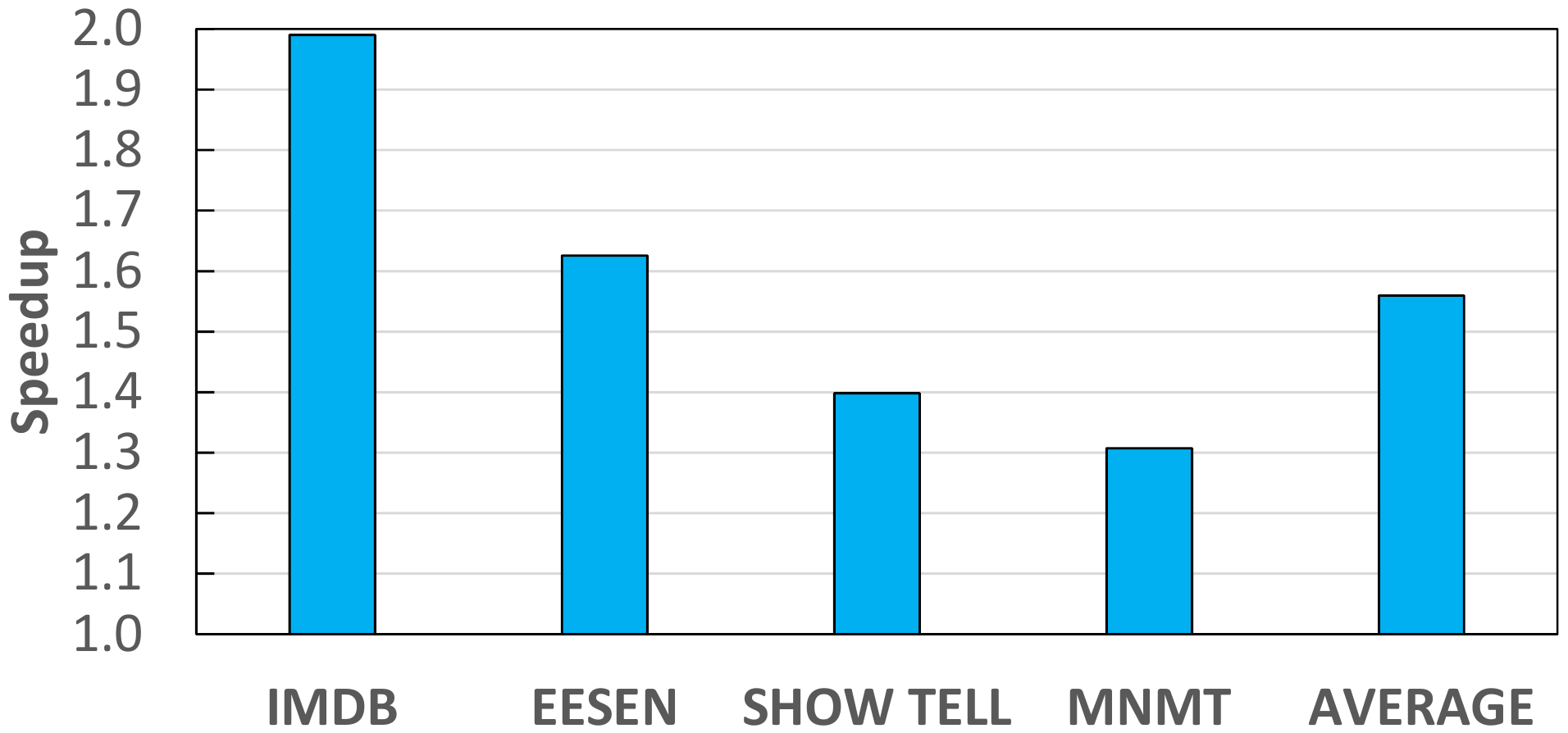}
	\caption{Speedups achieved by changing the precision dynamically. Baseline configuration is E-PUR+SIP.}
	\label{f:scheme_speedup}
\end{figure}

This section presents the evaluation of the proposed technique to dynamically select the precision based on the stability of the LSTM cell state. The baseline system is labelled as E-PUR+SIP, whereas the system including our dynamic precision selection scheme is labelled as E-PUR+DYN. We found that E-PUR+SIP has similar performance and energy consumption as the original version of E-PUR~\cite{fsilfa2018} that employs parallel 8-bit multipliers and, henceforth, we omit results for original E-PUR for the sake of brevity. The rest of this section is organized as follows. First, we present an evaluation of the effectiveness of our scheme. Second, we provide the performance and energy results. Finally, we analyze the area overheads of the proposed scheme.

Figure~\ref{f:random_accuracy_loss} reports the effectiveness of using the cell state stability in order to dynamically set the precision. In this figure, we compare our proposal with a scheme that randomly selects the precision level for each element of the cell state at each time-step. The random scheme has a low precision usage of 33\% on average, i.e. 33\% of the evaluations are performed at low precision (4 bits) whereas 67\% are done at high precision (8 bits). However, the random scheme produces a significant loss in accuracy for all the networks. On the other hand, our scheme has a 67\% low precision usage on average, without any loss in accuracy. Therefore, tracking the stability of the LSTM cell state provides valuable information to select the precision. The effectiveness of this approach is illustrated in Figure~\ref{f:cell_state_peak_fix}. As it can be seen, by evaluating the peaks of the cell state in high precision and the stable regions in low precision, our system accurately tracks the behavior of the 32-bit fp version. On average, we measured that the error in the peaks is reduced from 78\% to 26\%. Therefore, our scheme maintains accuracy while using low precision for a large percentage of the evaluations.

Figure~\ref{f:scheme_speedup} shows the performance improvements for our set of LSTM networks. On average, a speedup of 1.56x is obtained without any accuracy loss. 

As seen in Figure~\ref{f:scheme_speedup}, all the models achieve consistent and significant speedups when compared with the baseline. The reduction in execution time is due to using lower precision (4-bit) for more than 60\% of the time. Note that the baseline employs 8 bits to maintain the accuracy. Furthermore, the smaller the bit width the higher the performance of the SIP units: switching from 8 bits to 4 bits doubles the performance of the dot product units. Hence, for time-steps and cell state elements where 4-bit precision is used (stable phases), the latency of the dot product is reduced by a factor of 2x with respect to the 8-bit version. This represents around 60\% of the evaluations for our set of LSTM networks on average. On the other hand, 40\% of the evaluations are still done using 8 bits to maintain accuracy (peaks of cell state), and no performance improvement is achieved for those evaluations.

Note that the obtained is speedup is close to the theoretical speedup because the latency of our scheme is largely hidden, since its execution is overlapped with the computations in the multi-functional unit and all the SIP units are fully utilized. For all the networks, just a small execution time overhead is added when the evaluation of an LSTM cell starts (i.e., first neuron and first time step). The IMDB network has a speedup of 1.99x since it can be evaluate entirely in low precision, and our scheme is able to detect it. For IMDB no peaks are observed in the cell state and, hence, low precision is used for all the evaluations. The networks EESEN, NMT and SHOWTELL achieve 1.63x, 1.40x and 1.31x speedup respectively, since they require high precision for some of their evaluations.

\begin{figure}[t!]
	\centering
	\includegraphics[width=3.375in]{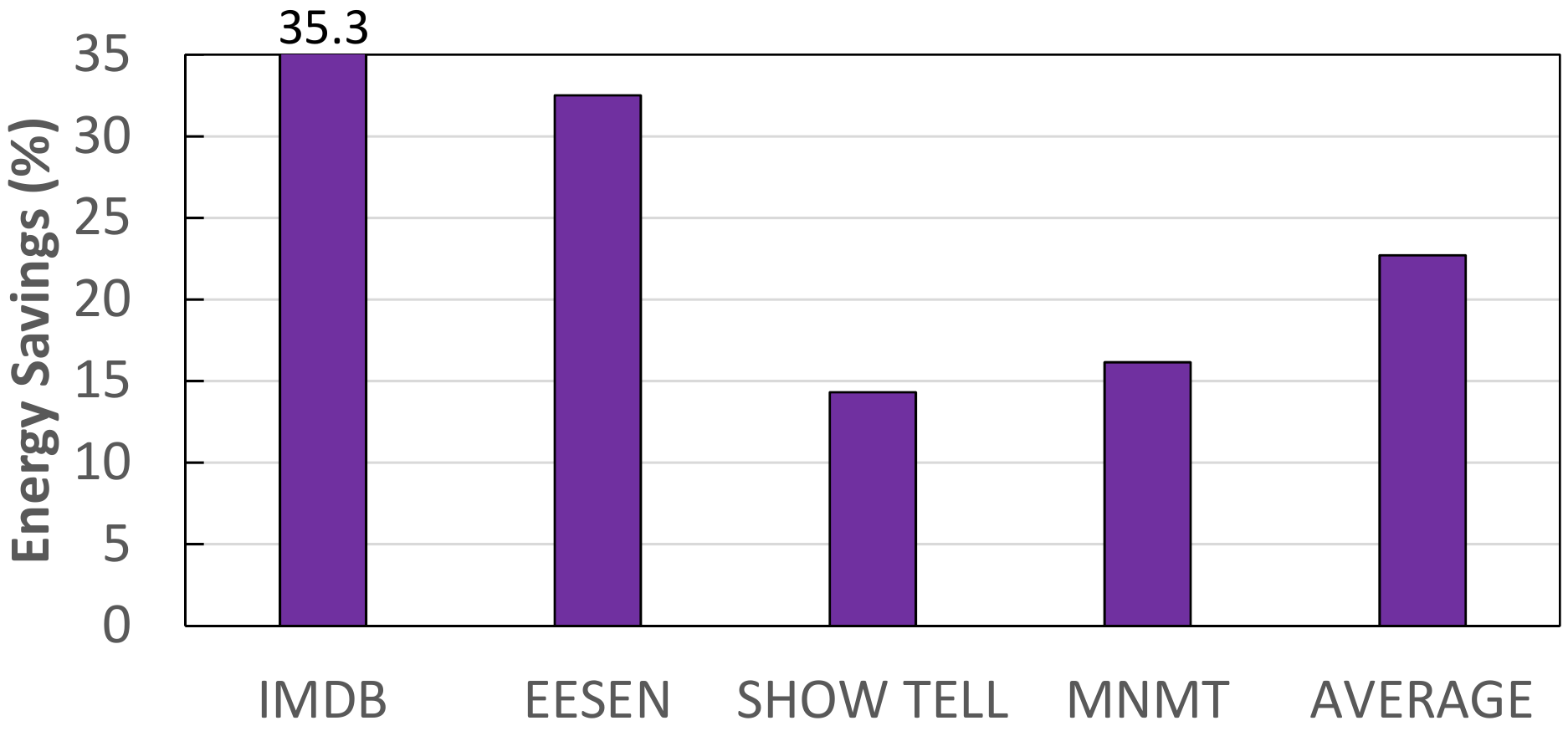}
	\caption{Energy savings achieved by dynamically changing the precision with respect to the baseline E-PUR+SIP.}
	\label{f:energy_savings}
\end{figure}

Figure~\ref{f:energy_savings} shows the energy savings achieved by E-PUR+DYN when compared with the baseline, including both static and dynamic energy. On average, E-PUR+DYN provides 23\% reduction in energy consumption. These savings come from several sources. First, using lower precision (4 bits) reduces the dynamic energy of the weight buffer, since less amount of information is fetched with respect to the 8-bit version. Second, the energy cost of the dot product is reduced when switching from 8 bits to 4 bits, since the activity in the SIP units is reduced. Finally, 
the speedups reported in Figure~\ref{f:scheme_speedup} provide a reduction in static energy. The LSTM networks EESEN and IMDB exhibit the largest energy savings, 32.5\% and 35.3\% respectively. For these networks, a large percentage of the computations are evaluated using low precision and, thus, the energy savings are significant. Most of the energy savings for these two networks are due to the reduction in static and dynamic energy of the scratchpad memories used to store the weights. For the networks SHOWTELL and MNMT, the energy savings are 14.3\% and 16.2\% respectively. These two are the largest models in our set of benchmarks and, hence, the overhead of the additional storage for the \textit{offset bit} has a larger impact on energy consumption than in the other two networks.

Regarding the area, EPUR+SIP has an area of 32.2 $mm^2$, whereas EPUR+DYN has an area of 35.3 $mm^2$. Hence, a small overhead of 8.8\% is added by the extra hardware required in order to support setting the precision dynamically. More specifically, the increase in area is due to the addition of extra memory to store the \textit{offset bit} and the hardware to implement the Peak Detector Unit.

%% file: related_work.tex
\section{Related Work}\label{s:related_work}

The area of hardware acceleration for LSTM networks has attracted the attention of the architectural community in recent years. The Tensor Processing Unit (TPU)~\cite{jouppi2017TPU} is an ASIC that supports convolutional, fully-connected and LSTM neural networks, delivering performance per watt orders of magnitude higher than CPUs and GPUs. It achieves a performance of 92 TOps/s (8-bit) while dissipating 40 Watts. E-PUR~\cite{fsilfa2018} is a recent accelerator specialized in LSTM networks that exploits temporal locality to minimize memory bandwidth usage and power dissipation (less than one Watt), while achieving low latency and real-time LSTM inference. Both accelerators, TPU and E-PUR, employ a fixed precision of 8 bits for weights and inputs. Our proposal is different as it selects the precision dynamically at runtime, using 4 bits for more than 66\% of the time without any accuracy loss.

On the other hand, more flexible accelerators that support variable precision have been introduced in recent years. Stripes~\cite{Stripes2016} uses a Serial Inner Product (SIP) unit in which the bits are fed serially and the bit width of the operands can be changed online. Stripes accelerates convolutional and fully-connected networks, but it does not support LSTM networks that is the main focus of this work. BitFusion~\cite{8416871} is a bit-flexible accelerator that includes an array of bit-level processing elements that can be dynamically merged or split to match the bitwidth of individual DNN layers. Furthermore, BitFusion provides full support for LSTM networks. Although Stripes and BitFusion provide more flexibility, the precision for each layer of a neural network is determined offline and it is kept constant during inference, i.e. the same layer employs the same precision for all the time steps. In this paper, we show that higher performance can be achieved by dynamically selecting the accuracy based on the evolution of the LSTM cell state. Our scheme is able to change the precision for every element of the cell state and every time step, further improving performance and energy efficiency.

In addition to the ASIC-based solutions, FPGA-based accelerators for LSTM inference have also been proposed in recent years. Brainwave~\cite{fowers2018configurable} is a Neural Processing Unit that achieves an order of magnitude improvement in latency  and throughput over state-of-the-art GPUs on large LSTM networks. The Efficient Speech Engine (ESE)~\cite{han2017ese} exploits pruning and sparsity to improve the performance of LSTM networks on FPGAs. C-LSTM~\cite{wang2018c} leverages structured compression techniques which reduce the LSTM model size while eliminating the irregularities of computation and memory accesses. On the other hand, DeltaRNN~\cite{gao2018deltarnn} and the work in~\cite{riera2018computation} exploit temporal coherency of the LSTM data to reuse computations and avoid redundant memory accesses. Pruning, compression and computation reuse techniques are completely orthogonal to our scheme.

To the best of our knowledge, none of the previous FPGA-based or ASIC-based accelerators is able to dynamically select the precision at runtime, on every LSTM cell element and time step, as our system does. This is the first work that employs the evolution of the values of the cell state to select the precision online.

%% file: conclusions.tex
\section{Conclusions}\label{s:conclusions}

In this paper, we present a novel scheme to dynamically select the precision for LSTM computation at runtime. We observe that the values of the LSTM cell state determine the required precision level: time steps where the value changes rapidly (i.e. peaks) require higher precision to avoid large errors, whereas time steps where the value is relatively stable can be evaluated with lower precision. Based on this observation, we propose a novel scheme that monitors recent values of the LSTM cell state and selects the appropriate precision for the next time step. Unlike previous schemes that fix the precision for each DNN layer offline, our system is able to change the precision for every cell state element and every time step.
We evaluate our proposal on top of a state-of-the-art accelerator for LSTM inference by using four popular LSTM networks. The experimental results show that our scheme selects the lowest precision for more than 66\% of the time without any loss in accuracy, providing  1.56x speedup and 23\% energy savings on average. The extra hardware required for our technique is quite modest and it represents a small area overhead of 8.8\%.